\documentclass[aps,prx,twocolumn,amsmath,amssymb,superscriptaddress]{revtex4-1}
\usepackage{graphicx}
\usepackage{dcolumn}
\usepackage{bm}

\begin{document}

\title{Quantum fluctuations and isotope effects in ab initio descriptions of water}

\author{Lu Wang}
\affiliation{Department of Chemistry, Stanford University, 333 Campus Drive, Stanford, California 94305}
\author{Michele Ceriotti}
\email{michele.ceriotti@epfl.ch}
\affiliation{Laboratory of Computational Science and Modeling, {\'E}cole Polytechnique F{\'e}d{\'e}rale de Lausanne, 1015 Lausanne, Switzerland}
\author{Thomas E. Markland}
\email{tmarkland@stanford.edu}
\affiliation{Department of Chemistry, Stanford University, 333 Campus Drive, Stanford, California 94305}

\date{\today}

\begin{abstract}
Isotope substitution is extensively used to investigate the microscopic behavior of hydrogen bonded systems such as liquid water. The changes in structure and stability of these systems upon isotope substitution arise entirely from the quantum mechanical nature of the nuclei. Here we provide a fully {\it ab initio} determination of the isotope exchange free energy and fractionation ratio of hydrogen and deuterium in water treating exactly nuclear quantum effects and explicitly modeling the quantum nature of the electrons. This allows us to assess how quantum effects in water manifest as isotope effects, and unravel how the interplay between electronic exchange and correlation and nuclear quantum fluctuations determine the structure of the hydrogen bond in water.
\end{abstract}


\maketitle

\section{\label{sec:level1}Introduction}
Nuclear quantum effects play a crucial role in determining the structure and dynamics of water's hydrogen bond network.\cite{Kuharski1984,Wallqvist1985,bern-thir86arpc,Stern2001,chen+03prl,Miller2005,morr-car08prl,Paesani2009,Paesani2009a,naga+12prl,liu+13jpcc,ceri+13pnas} For example, the zero point energy in a typical oxygen-hydrogen (O-H) covalent bond is equivalent to a 2000 Kelvin raise in temperature along that coordinate, and is strongly modulated by changes in the chemical environment. However, the vast majority of molecular dynamics simulations are performed treating the nuclei as classical particles, thereby neglecting these effects.

Recent development of efficient algorithms and ever increasing computational power now allow simulations of liquid water that treat both the nuclei and electrons quantum mechanically.\cite{chen+03prl,morr-car08prl,ceri+13pnas} These studies have uncovered a series of interesting observations. For instance, simulations using density functional theory (DFT) observed that the inclusion of nuclear quantum effects shortens the hydrogen bonds and leads to a more structured liquid.\cite{chen+03prl} However, another study using the same functional observed the opposite trend.\cite{morr-car08prl} More recent work has also suggested that, upon including nuclear quantum effects, protons in liquids water undergo transient events in which they hop onto nearby oxygen atoms, in stark contrast to what is observed with classical nuclei.\cite{ceri+13pnas} The degree of this hydrogen delocalization between molecules depends on the DFT functional chosen and hence it is not clear how pronounced this effect would be with an exact description of the electronic structure of liquid water - something that is still far beyond reach of computational investigation. 

Evaluating the role of nuclear quantum fluctuations in water and other hydrogen bonded systems is complicated due to the existence of competing quantum effects \cite{Stern2001,chen+03prl,habe+09jcp}. Quantum effects lead to an extension of the O-H covalent bond allowing the hydrogen to be shared (delocalized) between the oxygen atoms of hydrogen bonded pairs of water molecules. This acts to strengthen the hydrogen bond causing a structuring of the liquid and slower dynamics. In contrast, quantum fluctuations also allow the hydrogen to spread in other directions, distorting and weakening the hydrogen bond. Which of these effects dominates is determined by the strength of the hydrogen bond. ``Strong'' hydrogen bonds are made stronger upon the inclusion of nuclear quantum effects while ``weak'' ones are made weaker.\cite{li+11pnas} Liquid water sits at a point where the strength of its hydrogen bonds is such that a significant amount of the quantum effects cancel.\cite{chen+03prl,habe+09jcp,li+11pnas,mark-bern12pnas,McKenzie2014} Indeed it has recently been shown that raising the temperature leads to inversion from quantum fluctuations acting to strengthen water's hydrogen bonds to weakening them \cite{mark-bern12pnas}. The ability to treat competing quantum effects is thus a sensitive probe on whether an electronic structure method is able to correctly describe the behavior of hydrogen bonded systems - particularly at large covalent bond distances that would be extremely rarely observed in a classical simulation.

In order to assess the accuracy of different electronic structure methods in describing the effect of nuclear quantum fluctuations in water, one needs an experimental reference that selectively depends on the quantum nature of the nuclei.  While many properties of water, such as its structure or density, have both classical and quantum contributions, some properties are zero in the absence of nuclear quantum effects. Equilibrium isotope fractionation of hydrogen and deuterium between liquid water and its vapor is one such metric that is directly related to the quantum kinetic energy differences between the isotopes.\cite{mark-bern12pnas,ceri-mark13jcp} In addition, recent deep inelastic neutron scattering experiments have enabled the acquisition of absolute quantum kinetic energies of H and D atoms in different phases.\cite{reit+04bjp,pant+08prl,Pietropaolo2009,giul+11prl,flam+12jcp} Both of these properties can be obtained exactly for a given description of the electronic structure of water using the path integral formalism.

Here, we perform simulations incorporating the quantum nature of the nuclei and electrons to assess the ability of electronic structure methods to correctly describe the interplay between electronic and nuclear fluctuations in hydrogen bond networks. By combining recently developed techniques we extract the isotope fractionation ratios and quantum kinetic energies in liquid water and show how the local hydrogen bonded geometries engender changes in the quantum kinetic energy. This allows us to assess the influence of dispersion and electronic exchange interactions on the balance between hydrogen bond structuring and distortion in liquid water.

\section{\label{sec:level1}Materials and methods}
\textit{Ab initio} path integral molecular dynamics (AI-PIMD) simulations of water in the liquid and gas phase were performed with the i-PI wrapper\cite{ceri+14cpc} for path integral evolution and the Quickstep module in the CP2K package\cite{vand-krac05cpc} for the electronic structure calculations. Each atom was represented with 6 beads using the path integral generalized Langevin equation (PIGLET) algorithm.\cite{ceri-mano12prl,GLE4MD} Simulations were performed in the canonical (NVT) ensemble at the experimental density for each temperature studied with a time step of 0.5 fs. Liquid simulations were performed for 50 ps and gas phase simulations for 250 ps for each functional. Electronic structure calculations were carried out using four exchange-correlation functionals: BLYP,\cite{beck88pra,lee+88prb} PBE,\cite{Perdew1996,Perdew1997} B3LYP\cite{beck93jcp} and PBE0\cite{adam-baro99jcp} using the Goedecker-Teter-Hutter pseudopotentials.\cite{goed+96prb} In addition, simulations including the D3 dispersion correction\cite{grim+10jcp} were also performed with the BLYP and B3LYP functionals, denoted as BLYP-D3 and B3LYP-D3 in the following. The double-zeta split-valence basis set was utilized with a cutoff of 300 Rydberg for the representation of the charge density. Simulations using the PBE functional showed no statistically significant difference in the reported properties upon using a larger triple-zeta split-valence basis or a 500 Rydberg cutoff. Liquid water simulations were performed in a supercell with periodic boundary conditions, containing 64 water molecules when using GGA functionals, and 32 water molecules when using the more expensive hybrid functionals. Simulations using the BLYP functional showed no statistically significant difference in the reported properties when performed with 32 molecules, reflecting the local nature of the quantum kinetic energy.\cite{ceri-mark13jcp} For the gas phase simulations, the Martyna-Tuckerman Poisson solver was used with a cubic box of length 10 \AA.\cite{Martyna1999} Simulations of the Partridge-Schwenke monomer potential energy surface\cite{Partridge1997} were performed with the i-PI wrapper using the same simulation procedure as the AI-PIMD simulations but with 3 ns of sampling.
 
To calculate the fractionation ratios we used the thermodynamic free energy perturbation (TD-FEP) path integral estimator.\cite{ceri-mark13jcp} Combined with an appropriate choice of the integration variable to smooth the free energy derivatives\cite{ceri-mark13jcp} this allowed us to evaluate the isotope effects in the liquid phase using only a single PIMD trajectory of pure H$_2$O. Simulations using BLYP and PBE gave differences within the reported error bars when a second point at HOD was included, confirming that TD-FEP can extrapolate from H to D using a single simulation in the liquid. We note that since PIGLET does not enforce the imaginary time correlations needed to guarantee accelerated convergence for the TD-FEP estimators this check is essential to ensure the reliability of the results. \cite{ceri-mark13jcp} In the gas phase we found it necessary to perform separate simulations for isolated H$_2$O and HOD to obtain the required accuracy.\cite{ceri-mark13jcp}

\section{\label{sec:level1}Results and discussions}

\subsection{\label{sec:level2}Hydrogen bond fluctuations in liquid water}

\begin{figure}[h!tbp]
\centering
\includegraphics[height=15cm]{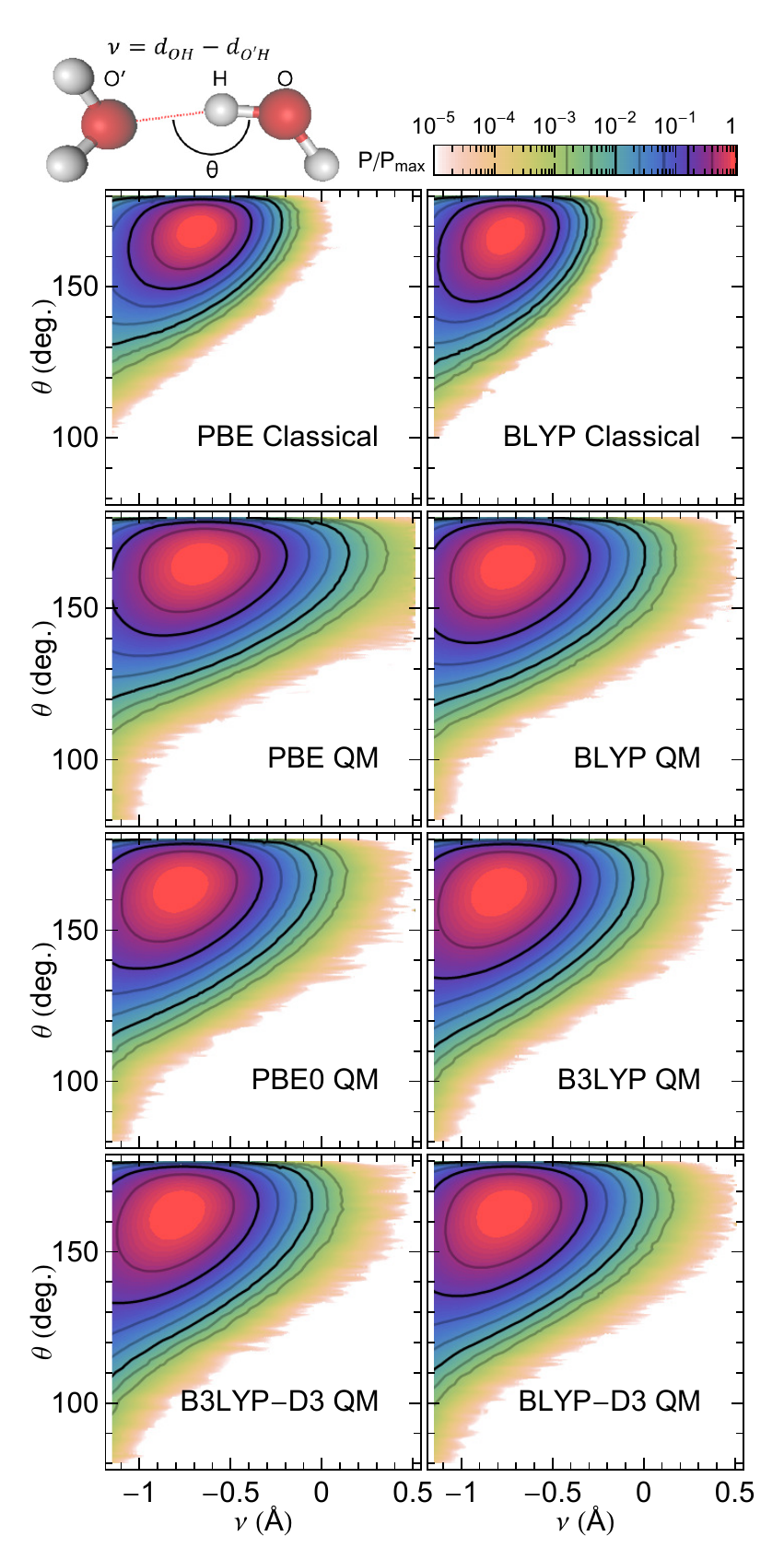}
\caption{ Probabilities of hydrogen bonded geometries sampled from AIMD simulations of liquid water as a function of the proton transfer coordinate $\nu$ and the OHO' hydrogen bond angle $\theta$ between a donor O and acceptor O' oxygen, as shown schematically at the top of the figure. $\nu=0$ corresponds to a proton which is exactly midway between O and O', while $\nu>0$ is associated with a proton that is undergoing a transient excursion to neighboring oxygen. $\theta=180^{\circ}$ corresponds to a perfectly linear hydrogen bond. The probability is normalized so that the maximum probability density is one. The nuclei are treated classically (Classical) or quantum mechanically (QM).}
\label{fig:RoomTemp}
\end{figure}

Fig. \ref{fig:RoomTemp} shows the probabilities of hydrogen bonded geometries obtained from our AI-PIMD simulations at 300~K using DFT descriptions of the electronic structure with classical and quantum treatment of the nuclei. The hydrogen bond is formed between a donor oxygen, O, that the proton is covalently bound to, and an acceptor oxygen, O'. To characterize the hydrogen bond geometries we define two coordinates $\nu$ and $\theta$ which are shown schematically in Fig. \ref{fig:RoomTemp}. The proton transfer coordinate is defined as $\nu = d_\text{OH}-d_\text{O'H}$ where $d_\text{OH}$ and $d_\text{O'H}$ are distances of the proton from O and O' respectively. This measures the degree of proton sharing between water molecules with the value $\nu=0$ corresponding to a highly shared proton that is equally close to the donor and acceptor oxygen atom. The $\theta$ coordinate is the OHO' angle of the hydrogen bond between the donor and acceptor which is a measure of hydrogen bond distortion from perfect linearity ($\theta=180^{\circ}$). The probability distributions in $\nu$ and $\theta$ in Fig. 1 allow us to probe the influence of nuclear quantum fluctuations on the competition between proton sharing and hydrogen bond distortion in water.

Including nuclear quantum effects leads to a large increase in the fluctuations of both $\nu$ and $\theta$ (Fig. 1) compared to classical simulations. This increase in both proton sharing and hydrogen bond distortion, with the former acting to strengthen hydrogen bonds and the latter weakening them, is a manifestation of competing quantum effects.\cite{Stern2001,chen+03prl,habe+09jcp} We note that the changes upon including nuclear quantum effects are much greater than those from changing the density functional or including dispersion corrections since the zero-point energy in each O-H stretch is $\sim5$ kcal/mol of energy ($\sim10 k_BT$ at 300 K) - a much larger energy scale than the differences in hydrogen bond strengths obtained from commonly used density functionals. This highlights the importance of including nuclear quantum effects when assessing the behavior of protons in hydrogen bonded systems, particularly those with short hydrogen bonds. 

Consistent with another recent study, our classical simulations exhibit no probability ($<10^{-7}$) of obtaining configurations where a proton is transiently closer to the acceptor than donor oxygen ($\nu>0$), while the quantum simulations exhibit a significant number of such extreme fluctuations.\cite{ceri+13pnas} These configurations are accompanied by a large electronic rearrangement.\cite{ceri+13pnas} Such an observation is in stark contrast with the traditional picture of the hydrogen bond in water as simply a strong electrostatic interaction. One should thus be highly careful in drawing conclusions about such proton behavior from traditional empirical force fields \cite{ceri+13pnas,frit+14jctc}. This motivates the need to develop next generation potentials that can accurately reproduce the forces from electronic structure using training sets incorporating the large bond and angle extensions that are observed in simulations including nuclear quantum effects \cite{frit+14jctc,Babin2014}. Finally, it is important to note that these transient proton excursions to the acceptor oxygen, which arise due to nuclear quantum fluctuations, should not be interpreted as autoionization, which requires full separation of the proton and hydroxide species and is thus a much rarer event.

However, while all the descriptions of the electronic structure show proton excursions when nuclear quantum effects are included, the frequency of configurations with $\nu>0$ varies wildly, from 0.442\% of all protons in PBE simulations to 0.038\% in B3LYP ones (see Table~\ref{ta:hfrac300}). The increased proton sharing in PBE (a wider spread in $\nu$) is accompanied by significantly less distorted hydrogen bonds than other functionals (a narrower spread in $\theta$).

How can one assess the validity of these predictions? While classically the kinetic energy of a particle is invariant to its position, confinement of a quantum mechanical particle leads to an increase in its zero-point energy which raises the quantum contribution to its kinetic energy. Thus expansion of the proton down the hydrogen bonding direction, $\nu$, due to proton sharing leads to a decrease in the proton kinetic energy along that direction. Likewise hydrogen bond distortion leads to a decrease in the proton kinetic energy in that direction. The quantum kinetic energy can thus be used to assess the predictions of electronic structure approaches as to the influence of quantum fluctuations on the hydrogen bond geometries visited in water. 

\subsection{\label{sec:level2}Competing quantum effects on hydrogen fractionation}

\begin{table*}
\caption{The proton excursion probability, fractionation ratio 10$^3$ln$\alpha$ and the decomposition of the fractionation ratio into its components for each DFT functional at 300 K. The experimental fractionation ratio\cite{Horita1994} is also listed. The three coordinates for the decomposition are the O-H covalent bond vector (O-H), the vector orthogonal to the O-H direction in the plane of the water molecule (Plane) and the vector perpendicular to the molecular plane (Orthogonal). Since the three directions are mutually orthogonal, their sum is equal to the overall fractionation ratio. Error bars of the total fractionation ratio are $\pm$ 5.}
\label{ta:hfrac300}
\begin{tabular*}{\hsize}{@{\extracolsep{\fill}}cccccccc}
& PBE & BLYP & BLYP-D3 & PBE0 & B3LYP & B3LYP-D3 & Exp\\
\hline \hline
Proton excursion (\%)  &   0.442  & 0.098   & 0.073 & 0.054 & 0.038  & 0.040 &- \\
\hline 
O-H  &  -409   &  -292  & -272 & -241 & -199 & -205 & -   \\
Plane  &  114   &  104  & 90 & 99 & 88 & 87 & -   \\
Orthogonal  &  278   &  250  & 230 & 232 & 213 & 213 &-   \\
\hline
$10^3 \ln \alpha$  &  -17   &  62  & 48 & 90 & 102 & 95 & 73   \\
\hline \hline
\end{tabular*}
\end{table*}

The quantum kinetic energy difference between H and D in a given system or phase, $i$, is related to the free energy change upon exchanging H for D exactly by\cite{Vanicek2007,ceri-mark13jcp}
\begin{equation}
\Delta A_{i} = -\int^{m_\text{D}}_{m_\text{H}} d\mu \frac{\left<K_{i}(\mu)\right>}{\mu},
\label{eq:deltaA}
\end{equation}
where $K_{i}(\mu)$ is the kinetic energy of a hydrogen isotope of mass $\mu$ in the system $i$ and can be computed within the path integral formalism using the centroid virial estimator \cite{herm-bern82jcp,cao-bern89jcp}. In particular the H/D liquid-vapor fractionation ratio, $\alpha$, serves as a sensitive experimental probe of the change in the quantum kinetic energy of a particle upon moving from the vapor to liquid phase.\cite{mark-bern12pnas,ceri-mark13jcp} This ratio is related to the free energy change of the reaction,
\begin{equation}
H_{2}O_\text{(l)} + HOD_\text{(v)} \rightleftharpoons H_{2}O_\text{(v)} + HOD_\text{(l)}.
\end{equation}
by,
\begin{equation}
\alpha = e^{- (\Delta A_\text{l} - \Delta A_\text{v} )/k_{B}T}.
\label{eq:alpha}
\end{equation}
Here $\Delta A_\text{l}$ and $\Delta A_\text{v}$ correspond to the free energy of converting H to D in the liquid and vapor phases, respectively, given by Eq. \ref{eq:deltaA}. The fractionation ratio is typically expressed as $10^{3}$ln$\alpha$ and hence we will refer to this as the fractionation ratio in what follows. Using this definition, a positive value ($10^3 \ln \alpha>0$) corresponds to a preference for H to reside in the gas phase compared to D and a negative value for it to reside in the liquid phase. In the classical limit the average kinetic energy of the isotope will be $3 k_B T/2$ regardless of phase (classical equipartition) and hence $\Delta A_\text{l}=\Delta A_\text{v}$ giving rise to zero fractionation ($10^3 \ln \alpha=0$). Isotope fractionation thus depends purely on the quantum nature of the nuclei and probes the change in the quantum kinetic energy between the liquid, where water participates in hydrogen bonds, and the vapor, where it does not. Due to the vital importance of fractionation ratios as inputs to current climatic models, a number of detailed experiments have been performed to obtain accurate values of $\alpha$ over a wide range of temperatures.\cite{Horita1994}

Table~\ref{ta:hfrac300} shows the fractionation ratio obtained from our AI-PIMD simulations at 300 K. Since these simulations exactly include nuclear quantum fluctuations the only approximation is the DFT functional employed. The fractionation ratio is observed to fall as the number of proton excursions increases. This correlation occurs since proton excursions lead to a strengthening of hydrogen bonds in the quantum liquid resulting in a preference for the more quantum (lighter) H isotope to reside in the liquid where it can form stronger hydrogen bonds, in turn kicking out the less quantum D isotope to the vapor, thus reducing  $10^3 \ln \alpha$. The PBE functional, which has the highest percentage of observed proton excursions, has such a low fractionation ratio that it incorrectly predicts enhancement of H in the liquid ($10^3 \ln \alpha<0$) indicating that the amount of proton sharing with PBE is unphysical.

To uncover the origins of the differences in the fractionation ratio we can decompose the total fractionation ratio into three internal coordinates of each water molecule --  the O-H covalent bond vector, the orthogonal vector in the plane of the water molecule, and the vector perpendicular to the molecular plane. By projecting the quantum kinetic energy onto these internal coordinates we can assess how the different components of the quantum kinetic energy change between the liquid and vapor phase giving rise to fractionation (Table~\ref{ta:hfrac300}).\cite{mark-bern12pnas} For all DFT functionals the ability to delocalize the proton along the hydrogen bond in the liquid leads to a decrease in the quantum kinetic energy in the O-H direction (the proton is less confined). This yields a negative contribution to fractionation that is opposed by the other two directions (where the proton is confined by the hydrogen bond in contrast to essentially free rotation in the vapor phase). The O-H contribution is more negative for PBE than for any other functional, which is consistent with the large number of proton excursions. The internal coordinates chosen are orthogonal and hence their sum gives the overall fractionation ratio, allowing us to assess the percentage cancellation in the overall fractionation ratio,
\begin{equation}
\textstyle
100 \times  \left( 1 - \frac{|10^3 \ln \alpha|}{|10^3 \ln \alpha_\text{O-H}| + |10^3 \ln \alpha_\text{Plane}| + |10^3 \ln \alpha_\text{Orthogonal}|} \right).
\end{equation}
The cancellation ranges from 80\% for B3LYP to 98\% for PBE demonstrating that water exhibits an extremely large competition between quantum hydrogen bond delocalization and distortion. 

Table \ref{ta:hfrac300} shows that, with the exception of PBE, all the functionals we tested are qualitatively correct in predicting the enhancement of D in the liquid ($10^3 \ln \alpha>0$) reflecting the dominance of the hydrogen bond weakening effect (due to distortion of hydrogen bonds) over the strengthening effect (due to proton sharing) upon including nuclear quantum effects in water at 300 K. However, the functionals differ in their quantitative agreement with experiment. Among the functionals investigated, BLYP gives closest agreement with experiment. Including a fraction of exact exchange (the B3LYP functional) leads to a rise in the fractionation ratio. This highlights the tendency of BLYP and other GGA functionals to excessively delocalize electron density allowing protons to be shared more easily between water molecules. The propensity to delocalize electrons is decreased upon including exact exchange, resulting in a reduction of the excursions of protons along the hydrogen bond. This leads to an increase of the O-H component of fractionation and an overall positive fractionation ratio, i.e., a preference for H to reside in the gas phase. The effect on the overall fractionation is partially compensated by the fact that the weakened hydrogen bond formed when exact exchange is included leads to broader fluctuations in the other directions, reducing the magnitude of the in-plane and orthogonal components which favor H residing in the gas phase. The effect of including exact exchange in the PBE functional (PBE0) is much larger than that for BLYP, which is consistent with the fact that PBE0 contains a larger fraction of exact exchange relative to B3LYP (25\% vs 20\%). The effect of including dispersion corrections is smaller, and results in a decrease in fractionation for both BLYP and B3LYP. The opposite effects of exact exchange and dispersion corrections suggest that the good quantitative agreement of BLYP with experiments is probably fortuitous, arising from a cancellation of errors. A quantitative description of quantum effects requires an electronic structure method that includes both terms, such as B3LYP-D3. For systems where the large computational cost of hybrid functionals is prohibitive BLYP can qualitatively and semi-quantitatively capture the influence of nuclear quantum effects on hydrogen bonds, and does not over-estimate proton excursions as much as PBE. Overall, the strong correlation between fractionation and proton excursion events suggests that about 0.1\% of all protons in water at any given time are undergoing an excursion at 300 K. 

\subsection{\label{sec:level2}Temperature dependence}
\begin{figure}[h!tbp]
\centering
\includegraphics[height=7cm]{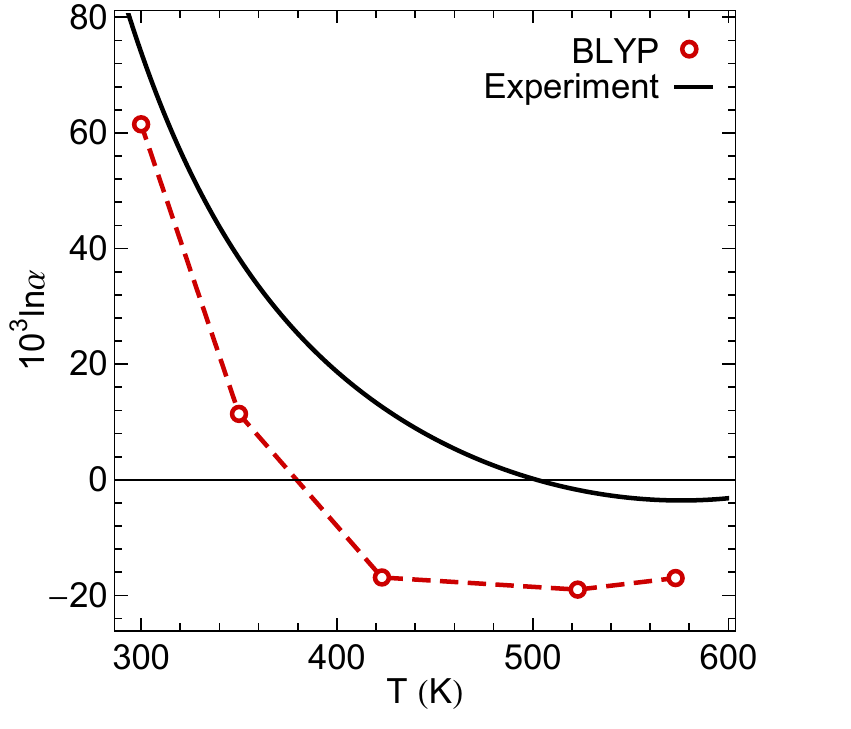}
\caption{Temperature dependence of the experimental\cite{Horita1994} and theoretical fractionation ratios 10$^3$ln$\alpha$ along the water liquid-vapor coexistence line. The theoretical points were obtained using the BLYP functional. The dashed line is a guide for the eye.}
\label{fig:HighT}
\end{figure}

Given BLYP's success in capturing qualitatively the interplay between nuclear quantum effects and hydrogen bonding at 300 K, it is important to assess if it can describe this subtle cancellation over a range of temperatures. Fig. \ref{fig:HighT} compares the fractionation ratio predictions of BLYP to experiment along the liquid-vapor coexistence line from 300 K to 573 K, which is near the critical point of water. BLYP is observed to correctly capture the experimental inversion around 500 K which is seen in H/D liquid-vapor fractionation of water but not in most other substances or for other isotopes.\cite{Horita1994} However, the inversion is overestimated with the entire fractionation curve shifted down by $10^3 \ln \alpha \sim 20$ over the entire temperature range studied. The temperature where inversion occurs is about 100 K lower than experiment, which is consistent with previous BLYP simulations using classical nuclei that found the liquid-vapor critical point to be about 100 K too low \cite{McGrath2005}. The inversion arises from the different effect of changing temperature on the three contributions to the fractionation shown in Table \ref{ta:hfrac300}.\cite{mark-bern12pnas} The O-H contribution corresponds to a high frequency stretching mode and is thus highly quantum mechanical even at the highest temperatures shown. In contrast the in-plane and orthogonal components, which oppose the O-H contribution to fractionation, correspond to lower frequency rotational/librational type modes. These modes become classical much faster and hence temperature acts to tune the different contributions: for example from our BLYP simulations, upon going from 300~K to 573~K these contributions drop by $91\%$ and $86\%$ respectively whereas the O-H contribution drops by only $79\%$. This coincides with a drop in the percentage of protons undergoing excursions by an order of magnitude between 300~K and 573~K highlighting the quantum, rather than thermal, origin of this effect.

\subsection{\label{sec:level2}Absolute quantum kinetic energy of protons in water}

\begin{table*}
\caption{Proton kinetic energies and the free energy differences between H and D in the gas and liquid phase at 300 K for each functional. The exact values in the gas phase are obtained from PIMD simulations of the Partridge-Schwenke monomer potential.\cite{Partridge1997} $\Delta A_\text{l}$ is obtained combining the experimental fractionation ratio\cite{Horita1994} and the exact $\Delta A_\text{v}$. For the kinetic energies, error bars are $\pm$ 0.20 meV and $\pm$ 0.07 meV for $T_v$ and $T_l$ respectively.  For the free energy differences, error bars are $\pm$ 0.10 meV and $\pm$ 0.03 meV for $\Delta A_\text{v}$ and $\Delta A_\text{l}$ respectively.}
\label{ta:tk_deltaA}
\begin{tabular*}{1\hsize}{@{\extracolsep{\fill}}cccccccc}
& PBE & BLYP & BLYP-D3 & PBE0 & B3LYP & B3LYP-D3 & Exact \\
\hline \hline
K$_{v}$ (meV)  &  147.1  & 145.6  & 145.8 & 152.3 & 149.5 & 149.5 & 151.1 \\
$\Delta A_\text{v}$ (meV)  &  -88.1 & -87.2   & -87.3 & -91.1 & -89.6  & -89.6 & -90.4 \\
\hline
K$_{l}$ (meV)  &   146.7  & 148.3  & 147.5 & 155.6 & 153.7 &  153.3 & - \\
$\Delta A_\text{l}$ (meV)  &   -87.7  & -88.8   & -88.5 & -93.4 & -92.3  & -92.1 & -92.3 \\
\hline \hline
\end{tabular*}
\end{table*}

While using fractionation allows one to assess the change in quantum kinetic energy upon going between the vapor and liquid phase, it is also instructive to consider the absolute quantum kinetic energy in each phase. This allows us to examine whether any of the functionals obtain the correct overall fractionation ratio from a fortuitous cancellation of errors between the phases. 

To obtain exact values of the kinetic energy (K) and free energy change ($\Delta A$) upon converting H$_{2}$O to HOD in the vapor, we performed PIMD simulations on the Partridge-Schwenke (PS) water monomer potential energy surface.\cite{Partridge1997} This surface was fit to high-level {\it ab initio} calculations and has been shown to reproduce the rovibrational line positions of water and its isotopomers up to 30,000~cm$^{-1}$ in energy to accuracies of 0.1~cm$^{-1}$ or better.\cite{Partridge1997} Path integral simulations on this potential therefore allow us to obtain values for the kinetic energy of H and D in gaseous water that are exact to the number of decimal places needed to distinguish between the DFT functionals descriptions (Table \ref{ta:tk_deltaA}). From this it is clear that the functionals incorporating exact exchange (PBE0 and B3LYP) improve on both the absolute prediction of the proton quantum kinetic energy ($K_{v}$) and the free energy difference between H and D  ($\Delta A_\text{v}$) in the gas phase. By combining the value of $\Delta A_\text{v}$ obtained from the spectroscopically accurate PS potential with the experimental fractionation ratio we can obtain a benchmark value for the H/D isotope free energy change in the liquid, $\Delta A_\text{l}$, shown in Table \ref{ta:tk_deltaA}. The poor performance of the PBE functional is again apparent as is the accuracy of the values yielded by PBE0 and B3LYP. The improved performance of PBE0 over PBE is consistent with previous comparisons of monomer deformation energies to high level electronic structure theory.\cite{Santra2009}

Finally we can compare the absolute proton kinetic energy in the liquid obtained from DFT with the value obtained from recent deep inelastic neutron scattering (DINS) experiments ($143\pm3$ meV).\cite{pant+08prl,Pietropaolo2009} This value is lower than predicted by any of the functionals and is closest to the PBE functional, which we have shown to incorrectly give inverted fractionation and also gives poor values for the H/D liquid and gas free energy change. B3LYP and PBE0 show the largest disagreement with the DINS result despite obtaining excellent agreement with all other properties. Unfortunately, the combination of experimental fractionation data and exact gas phase calculations does not currently allow us to quote an absolute reference for the quantum kinetic energy in the liquid to resolve this discrepancy. One possible solution is to use DINS experiments to measure the kinetic energy of D in mixtures of H$_{2}$O and D$_{2}$O and hence obtain a value for the quantum kinetic energy of D in the dilute limit. One could then use this to compute a value of $\Delta A_\text{l}$ that could be compared with our reference value, making it possible to assess the consistency of DINS and thermodynamic (fractionation) data.

\section{Conclusions}

We have shown how nuclear quantum effects give rise to large qualitative and quantitative changes in the local hydrogen bonded structure of water, such as the appearance of transient proton excursions to nearby oxygen nuclei. Accompanying these hydrogen bond rearrangements are changes in the underlying electronic structure, which in turn modulate the nuclear motion and the quantum kinetic energy of the nuclei. These effects manifest as isotope effects and can be understood and validated by experimental probes of the quantum kinetic energy. Our simulations, incorporating both nuclear and electronic quantum effects, have shown that the transient proton excursions are  overestimated in GGA DFT functionals, where their tendency to delocalize electron density leads to excess delocalization of the nuclei, a problem that is exacerbated upon including nuclear quantum effects. While BLYP manages to qualitatively capture the cancellation between competing quantum effects, PBE over-estimates dramatically the extent of quantum fluctuations of the proton and incorrectly predicts inverse fractionation at room temperature. 

Including exact exchange and adding dispersion corrections lead to much improved agreement both for the H/D isotope free energy changes within a given phase, and the fractionation ratios when isotopes are moved between phases. In particular functionals which incorporate both exact exchange and dispersion contributions, such as B3LYP-D3, provide an excellent choice for studying quantitatively isotope free energy changes in hydrogen bonded systems. This understanding will facilitate the tackling of problems where highly distorted hydrogen bonded structures occur, such as in the solvation of charge dense ions, proton and hydroxide defects and at aqueous interfaces, for which accurate empirical force fields are not typically available. Such an {\it ab initio} approach will thus enable enhanced elucidation and computational modeling of problems ranging from fractionation in atmospheric and biological systems, to isotope effects on reaction free energy barriers.

\begin{acknowledgments}
T.E.M. acknowledges support from a Terman fellowship, an Alfred P. Sloan Research Fellowship and Stanford University start-up funds. L.W. acknowledges a postdoctoral fellowship from the Stanford Center for Molecular Analysis and Design. This work used the Extreme Science and Engineering Discovery Environment (XSEDE), which is supported by National Science Foundation grant number ACI-1053575 (project number TG-CHE140013). M.C. acknowledges generous allocation of computer time from the Swiss National Supercomputing Centre (project id s466) and the Oxford Supercomputing Centre. The authors gratefully thank Tim Berkelbach and Aaron Kelly for a careful reading of the manuscript.
\end{acknowledgments}


\begin{thebibliography}{10}

\bibitem{Kuharski1984}
R.~A. Kuharski and P.~J. Rossky, ``Quantum mechanical contributions to the
  structure of liquid water,'' {\em Chemical Physics Letters}, vol.~103, no.~5,
  pp.~357 -- 362, 1984.

\bibitem{Wallqvist1985}
A.~Wallqvist and B.~Berne, ``Path-integral simulation of pure water,'' {\em
  Chemical Physics Letters}, vol.~117, no.~3, pp.~214 -- 219, 1985.

\bibitem{bern-thir86arpc}
B.~J. Berne and D.~Thirumalai, ``{On the Simulation of Quantum Systems: Path
  Integral Methods},'' {\em Annual Review of Physical Chemistry}, vol.~37,
  pp.~401--424, Oct. 1986.

\bibitem{Stern2001}
H.~A. Stern and B.~J. Berne, ``Quantum effects in liquid water: Path-integral
  simulations of a flexible and polarizable ab initio model,'' {\em The Journal
  of Chemical Physics}, vol.~115, no.~16, pp.~7622--7628, 2001.

\bibitem{chen+03prl}
B.~Chen, I.~Ivanov, M.~L. Klein, and M.~Parrinello, ``{Hydrogen Bonding in
  Water},'' {\em Phys. Rev. Lett.}, vol.~91, no.~21, p.~215503, 2003.

\bibitem{Miller2005}
T.~F. Miller and D.~E. Manolopoulos, ``Quantum diffusion in liquid water from
  ring polymer molecular dynamics,'' {\em The Journal of Chemical Physics},
  vol.~123, no.~15, p.~154504, 2005.

\bibitem{morr-car08prl}
J.~A. Morrone and R.~Car, ``{Nuclear Quantum Effects in Water},'' {\em Phys.
  Rev. Lett.}, vol.~101, no.~1, p.~17801, 2008.

\bibitem{Paesani2009}
F.~Paesani and G.~A. Voth, ``The properties of water: Insights from quantum
  simulationsâ€ ,'' {\em The Journal of Physical Chemistry B}, vol.~113,
  no.~17, pp.~5702--5719, 2009.

\bibitem{Paesani2009a}
F.~Paesani, S.~S. Xantheas, and G.~A. Voth, ``Infrared spectroscopy and
  hydrogen-bond dynamics of liquid water from centroid molecular dynamics with
  an ab initio-based force field,'' {\em The Journal of Physical Chemistry B},
  vol.~113, no.~39, pp.~13118--13130, 2009.

\bibitem{naga+12prl}
Y.~Nagata, R.~E. Pool, E.~H.~G. Backus, and M.~Bonn, ``{Nuclear Quantum Effects
  Affect Bond Orientation of Water at the Water-Vapor Interface},'' {\em Phys.
  Rev. Lett.}, vol.~109, no.~22, p.~226101, 2012.

\bibitem{liu+13jpcc}
J.~Liu, R.~S. Andino, C.~M. Miller, X.~Chen, D.~M. Wilkins, M.~Ceriotti, and
  D.~E. Manolopoulos, ``{A Surface-Specific Isotope Effect in Mixtures of Light
  and Heavy Water},'' {\em J. Phys. Chem. C}, vol.~117, pp.~2944--2951, Feb.
  2013.

\bibitem{ceri+13pnas}
M.~Ceriotti, J.~Cuny, M.~Parrinello, and D.~E. Manolopoulos, ``{Nuclear quantum
  effects and hydrogen bond fluctuations in water.},'' {\em Proc. Natl. Acad.
  Sci. USA}, vol.~110, pp.~15591--6, Sept. 2013.

\bibitem{habe+09jcp}
S.~Habershon, T.~E. Markland, and D.~E. Manolopoulos, ``{Competing quantum
  effects in the dynamics of a flexible water model.},'' {\em J. Chem. Phys.},
  vol.~131, p.~24501, July 2009.

\bibitem{li+11pnas}
X.-Z. Li, B.~Walker, and A.~Michaelides, ``{Quantum nature of the hydrogen
  bond},'' {\em Proc. Natl. Acad. Sci. USA}, vol.~108, pp.~6369--6373, Apr.
  2011.

\bibitem{mark-bern12pnas}
T.~E. Markland and B.~J. Berne, ``{Unraveling quantum mechanical effects in
  water using isotopic fractionation},'' {\em Proc. Natl. Acad. Sci. USA},
  vol.~109, no.~21, pp.~7988--7991, 2012.

\bibitem{McKenzie2014}
R.~H. McKenzie, C.~Bekker, B.~Athokpam, and S.~G. Ramesh, ``Effect of quantum
  nuclear motion on hydrogen bonding,'' {\em The Journal of Chemical Physics},
  vol.~140, no.~17, p.~174508, 2014.

\bibitem{ceri-mark13jcp}
M.~Ceriotti and T.~E. Markland, ``{Efficient methods and practical guidelines
  for simulating isotope effects.},'' {\em J. Chem. Phys.}, vol.~138,
  p.~014112, Jan. 2013.

\bibitem{reit+04bjp}
G.~Reiter, J.~C. Li, J.~Mayers, T.~Abdul-Redah, and P.~Platzman, ``{The proton
  momentum distribution in water and ice},'' {\em Braz. J. Phys.}, vol.~34,
  pp.~142--147, 2004.

\bibitem{pant+08prl}
C.~Pantalei, A.~Pietropaolo, R.~Senesi, S.~Imberti, C.~Andreani, J.~Mayers,
  C.~Burnham, and G.~Reiter, ``{Proton momentum distribution of liquid water
  from room temperature to the supercritical phase},'' {\em Phys. Rev. Lett.},
  vol.~100, no.~17, p.~177801, 2008.

\bibitem{Pietropaolo2009}
A.~Pietropaolo, R.~Senesi, C.~Andreani, and J.~Mayers, ``{Quantum effects in
  water: proton kinetic energy maxima in stable and supercooled liquid},'' {\em
  {Brazilian Journal of Physics}}, vol.~39, pp.~318 -- 321, 06 2009.

\bibitem{giul+11prl}
A.~Giuliani, F.~Bruni, M.~A. Ricci, and M.~A. Adams, ``{Isotope Quantum Effects
  on the Water Proton Mean Kinetic Energy},'' {\em Phys. Rev. Lett.}, vol.~106,
  no.~25, p.~255502, 2011.

\bibitem{flam+12jcp}
D.~Flammini, A.~Pietropaolo, R.~Senesi, C.~Andreani, F.~McBride, A.~Hodgson,
  M.~A. Adams, L.~Lin, and R.~Car, ``{Spherical momentum distribution of the
  protons in hexagonal ice from modeling of inelastic neutron scattering
  data.},'' {\em J. Chem. Phys.}, vol.~136, p.~024504, Jan. 2012.

\bibitem{ceri+14cpc}
M.~Ceriotti, J.~More, and D.~E. Manolopoulos, ``{i-PI: A Python interface for
  ab initio path integral molecular dynamics simulations},'' {\em Comp. Phys.
  Comm.}, vol.~185, pp.~1019--1026, Nov. 2014.

\bibitem{vand-krac05cpc}
J.~VandeVondele, M.~Krack, F.~Mohamed, M.~Parrinello, T.~Chassaing, and
  J.~Hutter, ``{Quickstep: Fast and accurate density functional calculations
  using a mixed Gaussian and plane waves approach},'' {\em Comp. Phys. Comm.},
  vol.~167, pp.~103--128, Apr. 2005.

\bibitem{ceri-mano12prl}
M.~Ceriotti and D.~E. Manolopoulos, ``{Efficient First-Principles Calculation
  of the Quantum Kinetic Energy and Momentum Distribution of Nuclei},'' {\em
  Phys. Rev. Lett.}, vol.~109, p.~100604, Sept. 2012.

\bibitem{GLE4MD}
M.~Ceriotti, ``{GLE4MD}.'' http://epfl-cosmo.github.io/gle4md/, 2010.

\bibitem{beck88pra}
A.~D. Becke, ``{Density-functional exchange-energy approximation with correct
  asymptotic behavior},'' {\em Phys. Rev. A}, vol.~38, no.~6, p.~3098, 1988.

\bibitem{lee+88prb}
C.~Lee, W.~Yang, and R.~G. Parr, ``{Development of the Colle-Salvetti
  correlation-energy formula into a functional of the electron density},'' {\em
  Phys. Rev. B}, vol.~37, no.~2, p.~785, 1988.

\bibitem{Perdew1996}
J.~P. Perdew, K.~Burke, and M.~Ernzerhof, ``Generalized gradient approximation
  made simple,'' {\em Phys. Rev. Lett.}, vol.~77, pp.~3865--3868, Oct 1996.

\bibitem{Perdew1997}
J.~P. Perdew, K.~Burke, and M.~Ernzerhof, ``Errata: Generalized gradient
  approximation made simple,'' {\em Phys. Rev. Lett.}, vol.~78, p.~1396, Feb
  1997.

\bibitem{beck93jcp}
A.~D. Becke, ``{Density-functional thermochemistry. III. The role of exact
  exchange},'' {\em J. Chem. Phys.}, vol.~98, no.~7, p.~5648, 1993.

\bibitem{adam-baro99jcp}
C.~Adamo and V.~Barone, ``{Toward reliable density functional methods without
  adjustable parameters: The PBE0 model},'' {\em J. Chem. Phys.}, vol.~110,
  no.~13, p.~6158, 1999.

\bibitem{goed+96prb}
S.~Goedecker, M.~Teter, and J.~Hutter, ``{Separable dual-space Gaussian
  pseudopotentials},'' {\em Phys. Rev. B}, vol.~54, no.~3, pp.~1703--1710,
  1996.

\bibitem{grim+10jcp}
S.~Grimme, J.~Antony, S.~Ehrlich, and H.~Krieg, ``{A consistent and accurate ab
  initio parametrization of density functional dispersion correction (DFT-D)
  for the 94 elements H-Pu.},'' {\em J. Chem. Phys.}, vol.~132, p.~154104, Apr.
  2010.

\bibitem{Martyna1999}
G.~J. Martyna and M.~E. Tuckerman, ``A reciprocal space based method for
  treating long range interactions in ab initio and force-field-based
  calculations in clusters,'' {\em The Journal of Chemical Physics}, vol.~110,
  no.~6, pp.~2810--2821, 1999.

\bibitem{Partridge1997}
H.~Partridge and D.~W. Schwenke, ``The determination of an accurate isotope
  dependent potential energy surface for water from extensive ab initio
  calculations and experimental data,'' {\em The Journal of Chemical Physics},
  vol.~106, no.~11, pp.~4618--4639, 1997.

\bibitem{frit+14jctc}
S.~Fritsch, R.~Potestio, D.~Donadio, and K.~Kremer, ``{Nuclear quantum effects
  in water: A multi-scale study},'' {\em J. Chem. Theory Comput.}, vol.~10,
  pp.~816--824, Jan. 2014.

\bibitem{Babin2014}
V.~Babin, G.~R. Medders, and F.~Paesani, ``Development of a first principles
  water potential with flexible monomers. ii: Trimer potential energy surface,
  third virial coefficient, and small clusters,'' {\em Journal of Chemical
  Theory and Computation}, vol.~10, no.~4, pp.~1599--1607, 2014.

\bibitem{Horita1994}
J.~Horita and D.~J. Wesolowski, ``Liquid-vapor fractionation of oxygen and
  hydrogen isotopes of water from the freezing to the critical temperature,''
  {\em Geochimica et Cosmochimica Acta}, vol.~58, no.~16, pp.~3425 -- 3437,
  1994.

\bibitem{Vanicek2007}
J.~Vanicek and W.~H. Miller, ``Efficient estimators for quantum instanton
  evaluation of the kinetic isotope effects: Application to the intramolecular
  hydrogen transfer in pentadiene,'' {\em The Journal of Chemical Physics},
  vol.~127, no.~11, p.~114309, 2007.

\bibitem{herm-bern82jcp}
M.~F. Herman, E.~J. Bruskin, and B.~J. Berne, ``{On path integral Monte Carlo
  simulations},'' {\em J. Chem. Phys.}, vol.~76, no.~10, p.~5150, 1982.

\bibitem{cao-bern89jcp}
J.~Cao and B.~J. Berne, ``{On energy estimators in path integral Monte Carlo
  simulations: Dependence of accuracy on algorithm},'' {\em J. Chem. Phys.},
  vol.~91, no.~10, p.~6359, 1989.

\bibitem{McGrath2005}
M.~J. McGrath, J.~I. Siepmann, I.-F.~W. Kuo, C.~J. Mundy, J.~VandeVondele,
  J.~Hutter, F.~Mohamed, and M.~Krack, ``Simulating fluid-phase equilibria of
  water from first principlesâ€ ,'' {\em J. Phys. Chem. A}, vol.~110,
  pp.~640--646, Sept. 2005.

\bibitem{Santra2009}
B.~Santra, A.~Michaelides, and M.~Scheffler, ``Coupled cluster benchmarks of
  water monomers and dimers extracted from density-functional theory liquid
  water: The importance of monomer deformations,'' {\em The Journal of Chemical
  Physics}, vol.~131, no.~12, p.~124509, 2009.

\end{thebibliography}
\end{document}